\documentclass[%
reprint,
superscriptaddress,
amsmath,amssymb,
aps,
prl,
]{revtex4-2}
\usepackage{subcaption}    
\usepackage{booktabs}
\usepackage{graphicx}
\usepackage{bm}
\usepackage[usenames,dvipsnames,table]{xcolor}
\usepackage[colorlinks=true,linkcolor=Blue,citecolor=Blue,urlcolor=Blue]{hyperref}
\usepackage[capitalise]{cleveref} 
\crefname{section}{Sec.}{Secs.}

\usepackage{mathtools}
\usepackage[justification=raggedright, singlelinecheck=false, font=small]{caption}

\usepackage{tikz}
\usetikzlibrary{shapes.geometric, arrows.meta, positioning}

\begin{document}


	\author{Iván Léon}
	\affiliation{Department of Applied Mathematics and Computer Science, Universidad de Cantabria, Santander, Spain}
	
	\author{Riccardo Muolo}
	\affiliation{RIKEN Center for Interdisciplinary Theoretical and Mathematical Sciences (iTHEMS), Saitama, Japan}
	\affiliation{Department of Systems and Control Engineering, Institute of Science Tokyo, Tokyo, Japan}
	
	\author{Yuanzhao Zhang}
	\affiliation{Santa Fe Institute, Santa Fe, New Mexico, USA}
	
	\author{Maxime Lucas}
	\affiliation{Department of Mathematics \& Namur Institute for Complex Systems (naXys), University of Namur, Namur, Belgium}
	\affiliation{Earth and Life Institute, UCLouvain, Belgium}

    \title{Symmetry-based selection rules for higher-order interactions in coupled oscillators}
    
	\date{\today}
	
	\begin{abstract}
		 Pairwise interactions among general nonlinear oscillators can be reduced, via phase reduction, to a Kuramoto-type phase coupling $\sin(- \theta_j+\theta_k )$.
		 For higher-order interactions, multiple phase couplings exist---such as $\sin(-2\theta_j+\theta_k+\theta_l )$ and $\sin(-\theta_j+2\theta_k-\theta_l)$.
         Since different nonpairwise coupling functions produce qualitatively different dynamics, it is important to understand which phase couplings should be included in coupled phase oscillator models.
         In this Letter, we establish selection rules for higher-order phase coupling functions.
         These selection rules, which can be applied without the need of explicit phase reduction, are solely based on the symmetry of the isolated oscillator velocity field and the $n$-body interaction functions.
	As phase reduction established the mechanistic basis for the Kuramoto model, our results provide a theoretical link between physical systems and higher-order phase models.
	\end{abstract}
	
	\maketitle


Synchronization is ubiquitous in nature, from the collective flashing of fireflies to the coherent firing of neurons~\cite{pikovskysynchronization,arenas2008}. The Kuramoto model has emerged as the canonical framework for understanding such phenomena, describing pairwise coupled phase oscillators 
$\dot \theta_j = \omega_j + \kappa \sum_k \sin(- \theta_j + \theta_k)$~\cite{kuramoto1975selfentrainment,kuramoto1984chemical}. 
The model can be derived via phase reduction, a technique allowing us to approximate the dynamics through a single variable: the phase~$\theta$. The phase model captures the essential features of the transition to synchronization, which explains its remarkable success at capturing the dynamics of diverse systems of weakly-coupled nonlinear oscillators~\cite{nakao2016phase,pietras2019networka,monga2019phase,nakao2018phase,kuramoto2019concept}.

Many real systems involve oscillators interacting nonlinearly in groups of three or more, from neuronal assemblies with shared inhibition to multi-molecule chemical reactions~\cite{matheny2019exotic,battiston2020networks,bianconi2021higher,battiston2021physics,bick_explosive,bick2023higher,boccaletti2023structure,muolo2024turing,millan2025topology}.
Such $n$-body interacting systems can be described by
\begin{equation}
	\dot{\bm{X}}_j = \bm{F}(\bm{X}_j) + \kappa\sum_{\bm{k}} 
	A_{j\bm{k}}\,\bm{G}(\bm{X}_j, \{\bm{X}_{\bm{k}} \}),
	\label{eq:original_system_general}
\end{equation}
where $\bm{k}=(k_1,\ldots,k_{n-1})$. 
The Kuramoto model has been generalized to account for these higher-order interactions, but several nonequivalent forms exist~\cite{tanaka2011multistable,Skardal2020,millan2020explosive,nurisso2024unified,leon2025theory,battiston2026collective}.
Already for $n=3$ (three-body interactions), two coupling functions arise, namely,
\begin{equation*}
	\sin(-2\theta_j+\theta_k+\theta_l) \quad \text{and} \quad 
	\sin(-\theta_j + 2\theta_k-\theta_l),
\end{equation*}
which are fundamentally different with respect to the harmonic of the focus oscillator $j$, which controls many aspects of the dynamics, such as cluster formation, linear stability, and invariant manifolds~\cite{bick16,stankovski2017coupling,gong2019lowdimensional,Skardal2020,lucas2020multiorder,wang2024coexistence,leon2024,skardal2025mixed}.
However, many recent studies on Kuramoto oscillators with higher-order interactions simply assume one or the other coupling function, mostly without justification~\cite{adhikari2023synchronization,huh2024critical,costa2024bifurcations,wang2026moderate}.


Phase reduction can produce both three-body couplings---either from nonpairwise coupled nonlinear oscillators as in~\cref{eq:original_system_general}~\cite{leon2025theory} or as second-order corrections to pairwise coupled nonlinear oscillators~\cite{leon2019phase,leon22a,nijholt2022emergent,mau2023highorder,bick2024higher}\footnote{Hamiltonian embedding also naturally produces both coupling types~\cite{moriame2025hamiltonian}}. However, analytical results are challenging to obtain and only exist for a handful of systems---primarily Stuart-Landau oscillators~\cite{Win80} and van der Pol oscillators~\cite{Leon23a}, where symmetry 
facilitates calculations.
As a consequence, we lack a general principle linking the structure of the original nonlinear system to the coupling functions in phase-reduced models, leaving us unable to choose coupling functions on principled grounds or determine which physical systems a given phase model represents.

In this Letter, we uncover a symmetry principle governing the structure of higher-order phase interactions generated by weakly-coupled nonlinear oscillators.
Inversion symmetry of the isolated dynamics, $\bm{F}(-\bm{X})=-\bm{F}(\bm{X})$, 
acts as a selection rule suppressing entire classes of phase interactions, 
with the surviving couplings further determined by the parity of $\bm{G}$.
This holds for group interactions of any size $n$: the parity of the harmonic of the focus oscillator is fixed by the parity of $\bm{G}$ with respect to $\bm{X}_j$.
When the symmetry of the isolated dynamics, or that of the original coupling, is broken, a mixture of all phase interactions is generally present. 
We prove these rules analytically via Fourier parity constraints, validate 
them across five representative oscillators spanning neural, chemical, and 
generic oscillatory systems, and demonstrate continuous tuning of the coupling type 
through controlled symmetry breaking in a modified Stuart--Landau oscillator.

\begin{figure}[t]
    \centering
    \includegraphics[width=0.99\linewidth]{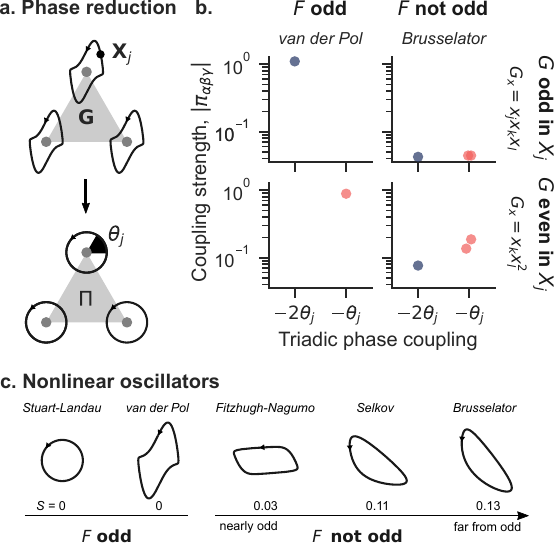}

    \caption{
    	\textbf{Symmetry of nonlinear oscillators suppresses classes of coupling in phase reduced models.}
    	\textbf{a.}~Schematic illustration of the phase reduction of nonlinear oscillators with three-body interactions. 
    	\textbf{b.}~We show the phase coupling coefficients $|\pi_{-2, 1, 1}|$ (blue), and $|\pi_{-1, 2, -1}|$ and $|\pi_{-1, -1, 2}|$ (red) for two representative oscillators: van der Pol (odd $\bm{F}$) and the Brusselator (non-odd $\bm{F}$).
    	\textbf{c.}~Limit-cycle of the five nonlinear oscillators, from purely odd to far from odd, with associated symmetry-breaking index $\mathcal{S}$.
    }
    \label{fig:numerical_classification}
\end{figure}

    
\textit{Phase reduction of higher-order interactions.---}%
We consider $N$ nonlinear oscillators with three-body interactions that evolve according to
\begin{equation}
    \dot{\bm{X}}_j=\bm{F}(\bm{X}_j)+\kappa\sum_{k,l} A_{jkl} \, \bm{G}(\bm{X}_j,\bm{X}_k,\bm{X}_l) ,
    \label{eq:original_system}
\end{equation}
where $\bm{X}_j \in \mathbb{R}^d$ is the $d$-dimensional state of oscillator $j$, $\bm{F}$ determines the isolated dynamics, $\bm{G}$ is a smooth coupling function, and the adjacency tensor $A_{jkl}$ specifies which triplets interact with strength $\kappa$. 
For the sake of clarity, we consider three-body interactions and identical oscillators---the results are valid for any $n$-body interactions and nearly-identical oscillators
\footnote{Throughout the Letter we consider only identical oscillators. Nevertheless, if oscillators are similar $\bm{F}_j=\bm{F}+O(\kappa)$, the results of the paper are valid, since the difference gives corrections of order $O(\kappa^2)$ to Eq.~\eqref{eq:phase_reduction}. If oscillators are not similar, the averaging procedure (see SM for details) cannot be performed and Eq.~\eqref{eq:pi_definition} becomes $\Pi(\theta_j,\theta_k,\theta_l)=\bm{Z}(\theta_j)\cdot \bm{P}(\theta_j,\theta_k,\theta_l)$. In our theory, this means terms $\pi_{\alpha\beta\gamma}$ with $\alpha+\beta+\gamma\not=0$ can emerge.}%
.
For $\kappa=0$, each oscillator admits a stable limit cycle $\bm{X}^{\rm c}(t)$ of period $T$, which can be re-parametrized as $\bm{X}^{\rm c}(\theta)$ through a phase $\theta$ that grows uniformly with frequency $\omega = 2 \pi / T$.
%
Under weak coupling, $\kappa \ll 1$, standard phase reduction theory~\cite{kuramoto1984chemical,nakao2016phase} states that each oscillator can be described by its phase $\theta_j \in \mathbb{S}^1$ and that \cref{eq:original_system} is approximated by 
\begin{equation} \label{eq:phase_reduction}
    \dot{\theta}_j=\omega+\kappa\sum_{kl} A_{jkl} \, \Pi(\theta_j,\theta_k,\theta_l)+O(\kappa^2) ,
\end{equation}
as illustrated in \cref{fig:numerical_classification}a. 
Here, the effective phase interaction is defined as 
\begin{equation}\label{eq:pi_definition}
   \Pi(\theta_j,\theta_k,\theta_l)=\frac{1}{2\pi}\int_0^{2\pi}\bm{Z}(\theta_j+\varphi)\cdot \bm{P}(\theta_j+\varphi,\theta_k+\varphi,\theta_l+\varphi)d\varphi,
\end{equation}
where the phase sensitivity function $\bm{Z}(\theta)$ measures 
how the system responds to perturbation at a given phase
and the function 
\begin{equation} \label{eq:p_definition}
    \bm{P}(\theta_j,\theta_k,\theta_l)=\bm{G}(\bm{X}_j^{\rm c}(\theta_j),\bm{X}_k^{\rm c}(\theta_k),\bm{X}_l^{\rm c}(\theta_l))
\end{equation}
is the original coupling function evaluated on the limit cycle. The integral in Eq.~\eqref{eq:pi_definition} averages out fast-oscillating terms of order $O(\kappa^2)$ \cite{Sanders2007Averaging}. 

Can we predict the structure of the phase coupling $\Pi(\theta_j,\theta_k,\theta_l)$ without going through the full phase reduction procedure?
From \cref{eq:pi_definition,eq:p_definition}, we see that $\Pi$ inherits its structure from the two ingredients that specify the physical system: (i) the isolated dynamics $\bm{F}$, which determines the limit cycle $\bm{X}^{\rm c}$ and phase sensitivity $\bm{Z}$, and (ii) the coupling function $\bm{G}$, which enters only through its projection $\bm{P}$ onto the cycle:
\begin{equation}\label{eq:logical_flow}
\begin{array}{cccccc}
\bm{F} &\longrightarrow& \bm{X}^{\rm c} &\longrightarrow& \bm{Z} & \\
       &                &          &        \searrow      &  & \\
       & &    \bm{G}       &\longrightarrow& \bm{P}
\end{array}
\left.\vphantom{\begin{array}{c}\bm{Z} \\ \bm{P}\end{array}}\right\}
\;\longrightarrow\;
\Pi . 
\end{equation}
This logic flow will guide us in predicting phase interactions from constraints on $\bm{F}$ and $\bm{G}$.

To study its structure, we express $\Pi$ in Fourier modes
\begin{equation}\label{eq:pi_fourier}
    \Pi(\theta_j,\theta_k,\theta_l)=\sum_{\substack{\alpha, \beta , \gamma=-\infty\\ \alpha+\beta+\gamma=0}}^\infty \pi_{\alpha  \beta \gamma}e^{\mathrm{i} (\alpha \theta_j +  \beta \theta_k + \gamma \theta_l)} ,
\end{equation}
where 
$\pi_{\alpha \beta \gamma}=\pi_{-\alpha, -\beta, -\gamma}^*$
to ensure that $\Pi(\theta_j,\theta_k,\theta_l)$ is real, and the resonance condition
\begin{equation}\label{eq:resonance}
	\alpha+\beta+\gamma=0
\end{equation}
is imposed by the integral in Eq.~\eqref{eq:pi_definition} \footnote{If the oscillators are not identical nor similar, the resonance condition is not fulfilled and other harmonic/interaction may emerge.}:
since 
$\theta_j\simeq\omega t$, a non-resonant term 
$e^{i(\alpha+\beta+\gamma)\omega t}$ oscillates rapidly and 
vanishes upon averaging~\cite{Sanders2007Averaging}.
Here, we are interested in the coefficients $\pi_{\alpha \beta \gamma}$, which determine the presence and strength of the associated coupling functions $\sin(\alpha \theta_j + \beta \theta_k + \gamma \theta_l)$ in $\Pi$.
We refer to $\alpha$, $\beta$,  and $\gamma$ as harmonics: values of 1 and 2 correspond to first and second harmonics, and so on. 
We focus in particular on the lowest-harmonic three-body couplings  $\pi_{-1, -1, 2}$ and $\pi_{-2, 1, 1}$, which correspond to the two canonical coupling functions heavily studied in the literature: $\sin(-\theta_j - \theta_k + 2 \theta_l)$ and $\sin(-2\theta_j + \theta_k+\theta_l)$, respectively~%
\footnote{These functions are sometimes referred to as symmetric and asymmetric, respectively, based on their invariance under the swapping of non-focus oscillators $k \leftrightarrow l$~\cite{battiston2026collective}.}.
We refer to them as first and second $j$-harmonic based on the value of 
$\alpha$, to emphasize the dynamical importance of these harmonics.
Harmonic decay makes $\pi_{\alpha \beta \gamma}$ coefficients with higher harmonics weaker and thus less relevant for the dynamics (see App.).

\begin{figure}[t]
	\centering
    \includegraphics[width=0.9\linewidth]{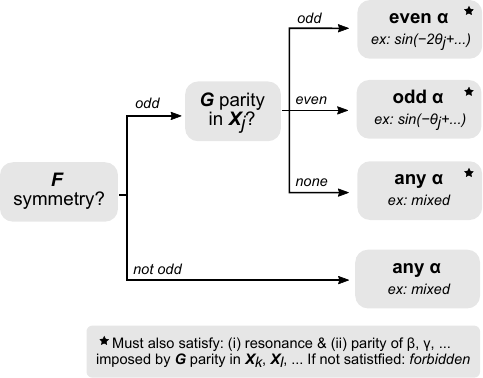}
    
	\caption{\textbf{Symmetry-based rules determining higher-order phase couplings.} Symmetry of $\bm{F}$ and parity of $\bm{G}$ in each argument $\bm{X}_m$ jointly determine the harmonics allowed in  $\sin(\alpha \theta_j + \beta \theta_k + \gamma \theta_l)$, subject to the resonance condition $\alpha+\beta+\gamma=0$. In particular, parity in the focus oscillator $\bm{X}_j$ determines the parity of $\alpha$ of the focus oscillator. }
	\label{fig:selection_rules}
\end{figure}

\textit{Numerical phase reduction reveals symmetry patterns.---}%
To uncover general patterns, we first numerically compute the phase reduction of five standard nonlinear oscillators $\bm{F}$---Stuart-Landau~\cite{kuramoto1984chemical,nakao2014complex}, van der Pol~\cite{vanderPol1926}, FitzHugh-Nagumo~\cite{fitzhugh,nagumo}, Selkov~\cite{Selkov1968,brechmann2018dynamics}, and Brusselator~\cite{PrigogineNicolis1967,PrigogineLefever1968}---with distinct dynamics, spanning neural, chemical, and generic oscillations (\cref{fig:numerical_classification}c), under twelve coupling functions $\bm{G}$ with different parity symmetries (see SM \cref{sec:app:oscillators}).

Two distinct behaviors emerge (\cref{fig:numerical_classification}b). For oscillators with odd isolated dynamics, $\bm{F}(-\bm{X})=-\bm{F}(\bm{X})$---such as Stuart-Landau and van der Pol---phase reduction produces only one of the two lowest-harmonic three-body phase couplings: either $-2\theta_j$ or $-\theta_j$ for the focus oscillator $j$, depending on the parity of $\bm{G}$. For oscillators without this odd symmetry---Selkov and Brusselator---both couplings are present simultaneously, regardless of $\bm{G}$ (see \cref{fig:sm:triadic_coefficients} for all oscillators and couplings). The ratio between the two coupling strengths is a consequence of the degree of symmetry breaking, which we measure with the symmetry-breaking index $\mathcal{S}(\bm{F}) \ge 0$ (see  App.).
The closer $\bm{F}$ is to being odd ($\mathcal{S} \simeq0$), the more one coupling dominates. As oddness breaks down ($\mathcal{S} > 0$), the ratio approaches unity. For example, for FitzHugh-Nagumo, which has a nearly odd $\bm{F}$, the ratio $\pi_{-1, -1, 2}/ \pi_{-2, 1, 1}$ is larger than that of the Brusselator, but smaller compared to Stuart-Landau (\cref{fig:sm:triadic_coefficients}). As we show next, this pattern is not coincidental but follows from exact analytical symmetry rules governing which phase interactions are allowed.

\textit{Analytical symmetry-based rules.---}%
We now show that this pattern follows from exact constraints on the Fourier coefficients $\pi_{\alpha\beta\gamma}$, which we analytically derive from the symmetries of $\bm{F}$ and $\bm{G}$ and summarized in the flow chart (\cref{fig:selection_rules}).
We start by expanding the limit cycle, phase sensitivity function, and coupling projection in Fourier modes
\begin{align}
    \bm{X}^{\rm c}(\theta) &= \sum_{n} \bm{x}^{\rm c}_n e^{\mathrm{i}n\theta}, \label{eq:xc_fourier}\\
    \bm{Z}(\theta) &= \sum_{m} \bm{z}_m e^{\mathrm{i}m\theta}, \label{eq:z_fourier} \\
    \bm{P}(\theta_j, \theta_k, \theta_l) &= \sum_{\alpha, \beta, \gamma} \bm{p}_{\alpha \beta \gamma} e^{\mathrm{i} (\alpha  \theta_j + \beta \theta_k + \gamma \theta_l)} . \label{eq:p_fourier} 
\end{align}
By injecting \cref{eq:xc_fourier,eq:z_fourier} into \cref{eq:pi_definition} and combining it with \cref{eq:pi_fourier}, the coefficients in $\Pi$ can  be expressed as 
\begin{equation}\label{eq:pi_coefficients}
    \pi_{\alpha \beta \gamma}=\sum_{\substack{m\\\alpha+\beta+\gamma=0}} \bm{z}_{\alpha-m} \cdot \bm{p}_{m\beta \gamma} ,
\end{equation}
which makes explicit how symmetries of $\bm{F}$---through $\bm{z}_{\alpha-m}$---and of $\bm{G}$---through $\bm{p}_{m\beta \gamma}$---jointly determine which $\pi_{\alpha\beta\gamma}$ can be nonzero, as illustrated in \cref{eq:logical_flow}.

\begin{figure}[t]
	\centering
	\includegraphics[width=3in]{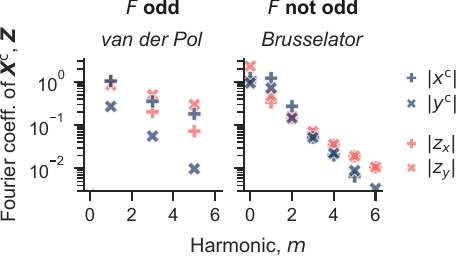}

	\caption{
		\textbf{Symmetry of nonlinear oscillators suppress even harmonics in limit cycle and phase sensitivity function.}	
		We show the modulus of each component of the Fourier coefficients $\bm{x}^{\rm c} = (x^{\rm c}, y^{\rm c})$ and $\bm{z} = (z_x, z_y)$ of $\bm{X^c}$ and $\bm{Z}$ [\cref{eq:xc_fourier,eq:z_fourier}], respectively, for a range of harmonics $m$,
        for the van der Pol and Brusselator oscillators. Even harmonics vanish in the van der Pol because of its inversion symmetry.}
	\label{fig:fourier_oscillator}
\end{figure}

%
Firstly, the symmetries of the vector field $\bm{F}$ impose conditions on the harmonics emerging on the limit cycle $\bm{X}^{\rm c}$ and phase sensitivity function $\bm{Z}$. If the isolated oscillator dynamics is odd
\begin{equation}
    \bm{F}(-\bm{X}) = -\bm{F}(\bm{X}),
\end{equation}
then both the limit cycle $\bm{X}^{\rm c}(\theta)$ and the phase sensitivity function $\bm{Z}(\theta)$ contain only odd harmonics: $\bm{x}^{\rm c}_{2n} =0$ and $\bm{z}_{2n}=0$~\cite{leon2025theory}.  
Indeed, oddness of $\bm{F}$ implies that the limit cycle must satisfy the $\pi$-shift symmetry $\bm{X}^{\rm c}(\theta+\pi) = -\bm{X}^{\rm c}(\theta)$, eliminating even harmonics---the same argument can be made for the phase sensitivity function $\bm{Z}$. We note that this condition forces $\alpha-m$, in \cref{eq:pi_coefficients}, to be odd when $\bm{F}$ is odd. If $\bm{F}$ is not odd, the limit cycle and phase sensitivity function can contain both even and odd harmonics and there is no such constraint. Numerics confirm this: for example, the van der Pol has odd $\bm{F}$ and only odd harmonics in $\bm{X}^{\rm c}$ and $\bm{Z}$, whereas the Brusselator is not odd and both even and odd harmonics are present (\cref{fig:fourier_oscillator}).
For oscillators that are only nearly odd, the even harmonics are substantially weaker than the odd ones, for example in the FitzHugh-Nagumo~(\cref{fig:sm:limit_cycle_coefficients}).
As we will show, the absence of even harmonics in $\bm{X}^{\rm c}$ and $\bm{Z}$, inherited from the oddness of $\bm{F}$, imply the suppression of entire classes of phase couplings in $\Pi$. 

Secondly, the parity of the coupling $\bm{G}$, combined with the symmetry of $\bm{F}$, imposes harmonic constraints on $\bm{P}$.
Namely, if $\bm{F}$ is odd and $\bm{G}$ is even ($+$) or odd ($-$) in any of its argument $\bm{X}_m$,
\begin{equation}\label{eq:parity_G}
\bm{G}(\dots, -\bm{X}_m, \dots)
= \pm \bm{G}(\dots,\bm{X}_m, \dots).
\end{equation}
then only even or odd, respectively, Fourier coefficients in $\theta_m$ survive in $\bm{P}$. 
This can be seen from \cref{eq:p_definition,eq:p_fourier}: it is caused by the $\pi$-shift symmetry propagating through $\bm{G}$ into $\bm{P}$ (see S.M. \cref{sec:sm:p_harmonics}).
If $\bm{G}$ is polynomial---as in most reaction–diffusion or population models---then it can be decomposed into the contributions of each of its monomials, which are odd (even) in $\bm{X}_m$ if $\bm{X}_m = (x_m, y_m)$ appears with odd (even) power.
For example, $\bm{G} = (x_j x_k x_l, 0)$---odd in all arguments---implies $\bm{p}_{\alpha\beta\gamma}\neq 0$ only for all odd $(\alpha,\beta,\gamma)$; whereas $\bm{G} = (x_kx_l^2,0)$---odd in $\bm{X}_k$, even in $\bm{X}_j$ and $\bm{X}_l$---forces odd $\beta$ but even $\alpha$ and $\gamma$. It is possible for $\bm{G}$ to have no parity in $\bm{X}$---non polynomial---then the corresponding coefficient is not constrained; similarly, if $\bm{F}$ is not odd, none of the coefficients are constrained.

%
Finally, we derive the allowed phase coupling coefficients $\pi_{\alpha \beta \gamma}$ by combining these constraints in~\cref{eq:pi_coefficients}.
If $\bm{F}$ is odd, $\alpha - m$ is odd, and the parity of $\alpha$ is determined by that of $m$---which we showed to be determined in $\bm{P}$ by the parity of $\bm{G}$ in $\bm{X}_j$. So, odd (even) $\bm{G}$ in $\bm{X}_j$ forces odd (even) $m$ and hence even (odd) $\alpha$, see \cref{fig:selection_rules}. Also, remember that the parity of $\beta$ and $\gamma$ is directly imposed by the parity of $\bm{G}$ in $\bm{X}_k$ and $\bm{X}_l$. 
The resonance condition in~\cref{eq:resonance} then acts as a compatibility check on the parities of the three coefficients: phase coupling satisfying all constraints are allowed; the others are suppressed (see S.M. \cref{sec:sm:pi_harmonics}).
For example, take an odd $\bm{F}$ and $\bm{G} = (x_j x_k x_l, 0)$ odd in all three arguments. Then, nonzero coefficients must have even $\alpha$ and odd $\beta$ and $\gamma$ in resonance---$\sin(-2\theta_j + \theta_k + \theta_l)$ is allowed, $\pi_{-2, 1, 1} \neq 0$, but $\sin(-\theta_j - \theta_k + 2 \theta_l)$ is not, $\pi_{-1, -1, 2}=0$. Conversely, with $\bm{G} = (x_k x_l^2, 0)$, we must have odd $\alpha$ and $\gamma$ but even $\beta$, and hence $\sin(-\theta_j - \theta_k + 2 \theta_l)$ is allowed but $\sin(-2\theta_j + \theta_k + \theta_l)$ is not. 
With $\bm{G} = (x_k x_l,0)$, the parity and resonance constraints are incompatible, and no phase coupling interactions are allowed.
When $\bm{F}$ is not odd, all parity constraints are lifted and both couplings are generically present regardless of $\bm{G}$. 
Importantly, since phase reduction is linear in $\bm{G}$, the rules apply term by term to polynomial couplings: for $\bm{G} = (x_jx_kx_l + x_kx_l^2, 0)$, the first term (odd in all arguments) selects $\sin(-2\theta_j+\theta_k+\theta_l)$ and the second (even in $\bm{X}_j$) selects $\sin(-\theta_j-\theta_k+2\theta_l)$, so both are present---consistent with the ``no parity in $\bm{X}_l$'' in \cref{fig:selection_rules}.
These rules are summarized in \cref{fig:selection_rules}. 

The described rules extend to $n$-body interactions of any size: the parity of $\alpha$ is always fixed by the parity of $\bm{G}$ in the focus oscillator $\bm{X}_j$, with remaining indices constrained by the parities of $\bm{G}$ in the other arguments and the resonance condition $\sum_i\alpha_i=0$ where $\alpha_i$ is the harmonic of the $i$-th phase (\cref{fig:selection_rules} and S.M. \cref{sec:app:nbody}).
These symmetry rules analytically determine which phase couplings are allowed and which are suppressed, without needing to explicitly perform phase reduction. 

Thanks to harmonic decay, we can obtain analytical approximations of these couplings' strengths, by truncating $\bm{X}^{\rm c}$ and $\bm{Z}$ to their two lowest harmonics, and injecting into \cref{eq:pi_coefficients} (see App. for details).
For $\bm{G} = (x_j x_k x_l,0)$ and odd $\bm{F}$, the leading contribution is
$\pi_{-2,1,1}^{\text{trun}} = x_1^2 x_1^\ast z_1^{x\ast}$ and $\pi_{-1,-1,2} = 0$,
%
confirming the symmetry rules: the forbidden coupling vanishes exactly, and the surviving one depends only on the first harmonic of $\bm{X}^{\rm c}$ and $\bm{Z}$. For non-odd $\bm{F}$, even harmonics contribute to both coefficients at the same order, explaining the mixed couplings of comparable magnitude in \cref{fig:numerical_classification}. The approximation is quantitatively accurate: for van~der~Pol ($\mu=2$), $|\pi_{-2,1,1}^{\text{trun}}|\approx 1.124$, within $0.1\%$ of the exact numerical value. Expressions for $\bm{G}=(x_kx_l^2,0)$ and all other harmonics are given in the SM.

\begin{figure}[h]
    \centering
    \includegraphics[width=0.99\linewidth]{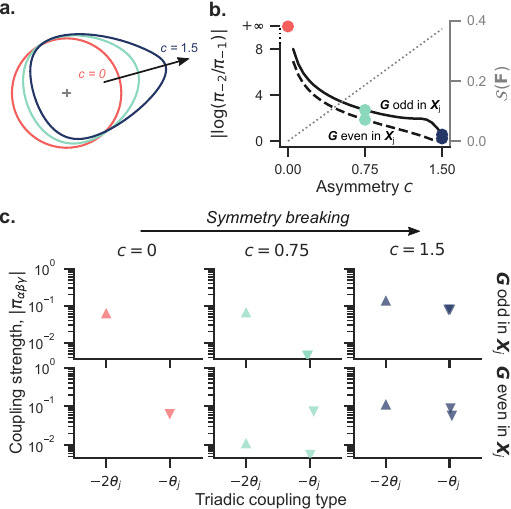}
    \caption{
    	\textbf{Controlled symmetry breaking tunes the three-body phase coupling type.}
    	\textbf{a.}~Limit cycles of the modified Stuart-Landau oscillator (Eq.~\ref{eq:design}) for increasing symmetry-breaking parameter $c$.
    	\textbf{b.}~Log-ratio of two types of coupling strengths $|\log(\pi_{-211}/\pi_{-1-12})|$ (solid/dashed lines) and symmetry breaking index $\mathcal{S}$ (dotted) as functions of $c$. Colored dots mark $c$ values shown in \textbf{a} and \textbf{c}.
    	\textbf{c.}~Three-body coupling strengths $|\pi_{\alpha\beta\gamma}|$ for the two coupling functions and three values of $c$. At $c=0$ only one coupling type is present per row, consistent with the parity rules; both grow to comparable magnitudes as symmetry breaking increases.
    }
    \label{fig:design}
\end{figure}

\textit{Design.---}%
The symmetry rules suggest a concrete design principle: 
to control the type of three-body phase coupling, one can tune the 
inversion symmetry of $\bm{F}$~\footnote{Following~\cite{namura2023designing}, one could go further and design an oscillator with the desired limit cycle and phase response function, allowing to control the form of the phase interactions.}. To demonstrate this, we modify the 
Stuart-Landau oscillator by adding an inversion-symmetry-breaking term 
controlled by $c$,
\begin{equation}\label{eq:design}
	\bm{F}_{\rm MSL}(\bm{X}) = \bm{F}_{\rm SL}(\bm{X}) + (cx^2,\, 0)^T,
\end{equation}
where $\bm{F}_{\rm SL}$ is the standard Stuart-Landau vector field. 
For $c=0$, $\bm{F}_{\rm MSL}$ satisfies $\bm{F}(-\bm{X})=-\bm{F}(\bm{X})$ 
exactly, and the symmetry rules predict a single phase coupling type depending 
on $\bm{G}$. For $c>0$, inversion symmetry is broken in a continuous and controlled way, introducing even harmonics into $\bm{X}^{\rm c}$ and $\bm{Z}$ 
with magnitude proportional to $c$ (\cref{fig:design}a).

Figure~\ref{fig:design}b-c confirms the predicted behavior. At $c=0$, 
only one coupling type is present for each $\bm{G}$: the second $j$-harmonic 
$\sin(-2\theta_j+\theta_k+\theta_l)$ for odd coupling $\bm{G} = (x_j x_k x_l, 0)$, 
and the first $j$-harmonic  $\sin(-\theta_j-\theta_k+2\theta_l)$ for even 
coupling  $\bm{G} = (x_k x_l^2, 0)$. As $c$ increases, the previously 
forbidden coupling grows from zero while the dominant one remains 
of comparable magnitude, and the ratio between them decreases 
monotonically toward unity. 


\textit{Discussion.---}
Our analytical symmetry rules establish which higher-order phase couplings emerge from weakly coupled nonlinear oscillators, and which do not, directly from the symmetries of $\bm{F}$ and $\bm{G}$, without requiring explicit phase reduction.
Given a phase model, the rules also admit a partial inverse at order $O(\kappa)$: exclusively even-$\alpha$ (odd-$\alpha$) coupling implies odd $\bm{F}$ with odd (even) $\bm{G}$ in $\bm{X}_j$; observing even and odd $\alpha$ implies broken inversion symmetry in $\bm{F}$ or mixed-parity $\bm{G}$. Note that a complete inverse is not possible, since multiple functions $\bm{F}$ and $\bm{G}$ provide identical phase interactions. 

The rules directly generalize to $n$-body interactions of any size: for odd $\bm{F}$, the parity of $\alpha$---the harmonic of the focus oscillator $j$---is fixed by the parity of $\bm{G}$ in $\bm{X}_j$, and the remaining harmonics are constrained by the parities of $\bm{G}$ in the other arguments and the resonance condition $\sum_i \alpha_i = 0$. 
Without symmetry of $\bm{F}$ or of $\bm{G}$ in $\bm{X_j}$, all phase coupling are generically allowed. 
The results also apply to oscillators of any dimension $d>2$ because they are solely based on symmetry arguments that are independent of oscillator dimension. 

These rules also shed new light on the well-studied pairwise case $n=2$. 
The resonance condition imposes $\beta=-\alpha$, so the coupling reduces to $\sin(\alpha(\theta_k-\theta_j))$\cite{kuramoto1984chemical}.
So, for most choices of $\bm{F}$ and $\bm{G}$, the Kuramoto-coupling $\sin(\theta_k-\theta_j)$ dominates, sometimes mixed with the weaker second harmonic $\sin(2\theta_k-2\theta_j)$\cite{CP16}. However, the Kuramoto interaction can be  completely suppressed if $\bm{F}$ and $\bm{G}$ are odd in $\bm{X_j}$, such as  $\bm{G} = (x_j x_k^2, 0)$. Thus, the selection rules provide (non-generic) conditions under which the oscillator's dynamics are not described by the Kuramoto model.

These results have a natural physical interpretation.
Odd $\bm{F}$ arises generically near any supercritical Hopf bifurcation---from rotational symmetry---or from physical inversion symmetry, as in van der Pol or mechanical oscillators \cite{matheny2014phase}. 
Non-odd $\bm{F}$ characterizes oscillators far from onset whose fixed point is displaced from the origin, as in circadian rhythms~\cite{goldbeter1995model} and conductance-based neurons~\cite{hodgkin1952quantitative}, generically activating both coupling types.
This classification provides principled guidance for choosing phase models without full phase reduction.

Three interesting directions remain open.
First, our results apply to systems with explicit higher-order coupling $\bm{G}$, not to higher-order terms arising as $O(\kappa^2)$ corrections to pairwise coupling~\cite{leon2019phase,leon22a,mau2023highorder}, where our rules may require modification.
Second, we focused on lowest-order harmonics justified by rapid harmonic decay, but higher harmonics may matter for strongly nonlinear oscillators.
Third, the resonance condition holds for nearly-identical oscillators---for more heterogeneous populations, non-resonant terms $\pi_{\alpha\beta\gamma}$ can emerge, allowing phase couplings forbidden by our rules.

As phase reduction established the mechanistic generality of the pairwise Kuramoto model, the symmetry rules derived here provide an analogous foundation for higher-order network dynamics---linking the structure of phase models directly to the symmetry class of the underlying nonlinear physical system.

\noindent \textbf{Acknowledgments:} 
The authors thank Timoteo Carletti, Marie Dorchain, and Rommel Tchinda Djeudjo for useful feedback on the manuscript. I.L. acknowledges support by MICIU/AEI/10.13039/50110001 1033
through Grant No. PID2021-125543NB-I00 and by ERDF/EU.
R.M. acknowledges JSPS KAKENHI 24KF0211 for financial support. 
Y.Z. acknowledges support by the Santa Fe Institute and the National Science Foundation under grant DMS-2436231.
M.L. is a Postdoctoral Researcher of the Fonds de la Recherche Scientifique–FNRS.

\noindent \textbf{Code availability:} 
Code for reproducing our results is available online from the repository \url{https://github.com/maximelucas/phase_reduction}.

\noindent \textbf{Author contributions:} 
Conceptualization: all authors.
Theoretical analysis: I.L.
Simulations: I.L. and R.M.
Visualization: M.L.
Writing -- original draft: I.L. and M.L.
All authors discussed the results, reviewed and edited the manuscript.

\noindent \textbf{Competing interests:} 
The authors declare no competing interests.

\bibliography{bib}

\clearpage

\section{Appendix}

\subsection{High harmonic decay}
\label{sec:app:harmonic_decay}

The amplitude $|\pi_{\alpha \beta \gamma}|$ decays rapidly with increasing harmonic order $|\alpha|+|\beta|+|\gamma|$, which can be quantified from standard Fourier regularity.  
If the limit cycle $\bm{X}^{\rm c}(\theta)$ is $C^p$, its Fourier coefficients decay algebraically, $|a_n|\!\sim\! n^{-p}$ for large $n$, or exponentially if $\bm{X}^{\rm c}$ is analytic~\cite{katznelson2004introduction}.  
Since the adjoint equation \cref{eq:adjoint} involves the same smooth periodic coefficients, $\bm{Z}$ inherits comparable Fourier decay, $|\bm{z}_m|\!\sim\! m^{-p}$.  
For polynomial or analytic couplings $\bm{G}$, the coefficients of $\bm{P}$ obey comparable bounds, and the convolution in \cref{eq:pi_coefficients} then yields $|\pi_{\alpha \beta \gamma}|\!\sim\!(|\alpha|+|\beta|+|\gamma|)^{-p}$ (or faster).  
Thus, the smoothness of the underlying limit cycle controls the spectral content of the phase coupling: higher harmonics are necessarily suppressed, explaining the numerical dominance of the lowest terms. We confirmed numerically that this decay already occurs at low $n$ in the five oscillators we considered (\cref{fig:harmonic_decay}).
Hence, although many combinations are in principle allowed, this spectral decay justifies focusing on the lowest-order three-body resonant terms: $\sin(\theta_k+\theta_l-2\theta_j)$ and $\sin(2\theta_k-\theta_l-\theta_j)$. 


\subsection{Inversion-symmetry breaking index}
\label{sec:app:symmetry_breaking}

To quantify the degree of inversion symmetry breaking of a vector 
field $\bm{F}$, that is, how far from satisfying oddness $\bm{F}(-\bm{X}) = -\bm{F}(\bm{X})$ it is, we define the inversion-symmetry breaking index
\begin{equation} \label{eq:symmetry_breaking_index}
    \mathcal{S}(\bm{F}) = \frac{\|\bm{F}_{\rm even}\|}{\|\bm{F}_{\rm odd}\|},
\end{equation}
where $\bm{F}_{\rm even}(\bm{X}) = \tfrac{1}{2}[\bm{F}(\bm{X})+\bm{F}(-\bm{X})]$ 
and $\bm{F}_{\rm odd}(\bm{X}) = \tfrac{1}{2}[\bm{F}(\bm{X})-\bm{F}(-\bm{X})]$ 
are the even and odd parts of $\bm{F}$, and $\|\cdot\|$ denotes the 
root-mean-square norm averaged over circles of varying radii in state space.
By construction, $\mathcal{S}=0$ if $\bm{F}$ is odd, 
and $\mathcal{S}$ increases monotonically with inversion symmetry breaking.
For polynomial vector fields, $\bm{F}_{\rm even}$ reduces to the 
sum of even-degree monomials; for the standard oscillators considered 
here, this is simply the constant term $\bm{F}(\bm{0})$, giving the 
fully analytic expression $\mathcal{S} = \|\bm{F}(\bm{0})\|/\|\bm{F}_{\rm odd}\|$.
For the modified Stuart--Landau oscillator~\eqref{eq:design}, 
$\bm{F}_{\rm even}(\bm{X}) = (cx^2, 0)^T$, which vanishes at the 
origin but grows with $c$, yielding $\mathcal{S}\propto |c|$ 
at any fixed sampling radius.

\subsection{Low-harmonic approximation}
\label{sec:app:low_harmonic}

To obtain analytical expressions for the coupling magnitudes, 
we truncate the limit cycle and phase sensitivity function to 
their two lowest harmonics,
\begin{align}
    X(\theta) &= x_0 + x_1 e^{i\theta} + x_2 e^{2i\theta} + \text{c.c.}, \\
    Z(\theta) &= z_0 + z_1 e^{i\theta} + z_2 e^{2i\theta} + \text{c.c.},
\end{align}
treating $X$ and $z$ as scalars for clarity (the vector case is 
analogous; see SM). The harmonic coefficients decay as 
$x_k, z_k \sim \epsilon^k$, so keeping terms to order $\epsilon^4$ 
captures the dominant contributions. Substituting into 
\cref{eq:pi_coefficients} for a given $\bm{G}$ and collecting 
resonant terms yields explicit expressions for all 
$\pi_{\alpha\beta\gamma}$.

For the two representative coupling functions considered in the 
main text, the lowest-order nonzero coefficients are as follows.
For $\bm{G}=(x_jx_kx_l,0)$:
\begin{align}
    \pi_{-2,1,1} &= \bigl(x_1^2 x_2^\ast z_0 
                  + x_1^2 x_1^\ast z_1^\ast 
                  + x_0 x_1^2 z_2^\ast\bigr)
                  + O(\epsilon^6), \\
    \pi_{-1,-1,2} &= \bigl(x_1^{\ast 2} x_2 z_0 
                   + x_0 x_1^\ast x_2 z_1^\ast\bigr)
                   + O(\epsilon^6),
\end{align}
with $\pi_{-1,-1,2}=\pi_{-1,2,-1}$, ensuring the full phase 
model preserves the permutation symmetry of $\bm{G}$.
For $\bm{G}=(x_kx_l^2,0)$:
\begin{align}
    \pi_{-2,1,1}   &= 2x_0 x_1^2 z_2^\ast\,
                    + O(\epsilon^6), \\
    \pi_{-1,-1,2}  &= \bigl(x_1^2 x_1^\ast z_1 
                    + 2x_0 x_1^\ast x_2 z_1^\ast\bigr) 
                    + O(\epsilon^6), \\
    \pi_{-1,2,-1}  &= 2x_0 x_1^\ast x_2 z_1^\ast\, 
                    + O(\epsilon^6).
\end{align}
For odd $\bm{F}$, even harmonics vanish 
($x_0=x_2=z_0=z_2=0$), and both sets reduce to a single  but different nonzero coefficient---$\pi_{-2,1,1} = x_1^2 x_1^\ast z_1^\ast$ and 
$\pi_{-1,-1,2} = x_1^2 x_1^\ast z_1$ respectively---consistent with 
the symmetry rules. 
For non-odd $\bm{F}$, all coefficients are allowed and are all of similar order $\sim\epsilon^4$, consistent with our observations too. 

Although we focused on purely three-body couplings with $\alpha, \beta, \gamma \neq 0$, this low-order approximation yields expressions with null coefficients too, including pairwise-like contributions at order $\epsilon^2$ 
when $x_0\neq 0$ and forcing-like triplets $(0,\beta,\gamma)$ 
present even for odd $\bm{F}$. Full tables of all nonzero 
coefficients for both coupling functions, and discussion of 
these additional terms, are given in the SM \cref{sec:sm:low-harmonics}.

\clearpage

\setcounter{figure}{0}
\setcounter{table}{0}
\setcounter{equation}{0}
\setcounter{page}{1}
\setcounter{section}{0}

\makeatletter
\renewcommand{\thefigure}{S\arabic{figure}}
\renewcommand{\theequation}{S\arabic{equation}}
\renewcommand{\thetable}{S\arabic{table}}

\clearpage

\setcounter{secnumdepth}{2} 
    
\widetext
\begin{center}
	\textbf{\large Supplementary Material: \\ Symmetry-based selection rules for higher-order interactions in coupled oscillators}

    \medskip
    Iván Léon, Riccardo Muolo, Yuanzhao Zhang, and Maxime Lucas
\end{center}


\section{Coupled oscillators models}
\label{sec:sm:models}

Here, we formally define the five oscillator models $\bm{F}$ and twelve coupling functions $\bm{G}$ that we used for the dynamics in \cref{eq:original_system}.

\subsection{Nonlinear oscillators}
\label{sec:app:oscillators}

We used five standard limit-cycle oscillator models, with qualitatively distinct dynamics spanning neural, chemical, and generic oscillations. They also represent different symmetry classes: Stuart-Landau~\cite{kuramoto1984chemical,nakao2014complex} and van der
Pol~\cite{vanderPol1926} have perfectly odd vector field, $\bm{F}(-\bm{X}) = -\bm{F}(\bm{X})$, whereas FitzHugh-Nagumo~\cite{fitzhugh,nagumo}, Selkov~\cite{Selkov1968,brechmann2018dynamics}, and Brusselator~\cite{PrigogineNicolis1967,PrigogineLefever1968} do not.  

\paragraph{Stuart-Landau oscillator.}
The Stuart-Landau oscillator is the normal form of a supercritical Hopf bifurcation and describes generic oscillations close to onset. Its vector field reads
\begin{equation}
	\bm F_{\rm SL}(x,y)
	=
	\begin{pmatrix}
		x-\alpha y-(x^2+y^2)(x-\beta y) \\
		y+\alpha x-(x^2+y^2)(y+\beta x)
	\end{pmatrix},
\end{equation}
with parameters $\alpha$ and $\beta$ controlling the natural frequency and nonisochronicity. In our simulations, we used $\alpha=1$ and $\beta=0$. 

\paragraph{van der Pol oscillator.}
The van der Pol oscillator is a paradigmatic nonlinear relaxation oscillator with amplitude-dependent damping. Its vector field is
\begin{equation}
	\bm F_{\rm vdP}(x,y)
	=
	\begin{pmatrix}
		y \\
		-x+\mu(1-x^2)y
	\end{pmatrix},
\end{equation}
with $\mu=2$. 

\paragraph{FitzHugh--Nagumo oscillator.}
The FitzHugh--Nagumo model is a prototypical reduction of conductance-based neuronal dynamics and captures excitable and oscillatory behavior in neurons. The vector field is
\begin{equation}
	\bm F_{\rm FHN}(x,y)
	=
	\begin{pmatrix}
		x-\dfrac{x^3}{3}-y \\
		\delta(x+a-b y)
	\end{pmatrix},
\end{equation}
with parameters $a=0.8$, $b=0.8$, and $\delta=0.08$. The term $\delta a$ slightly breaks the odd symmetry. 

\paragraph{Selkov oscillator.}
The Selkov model is a minimal two-variable model of glycolytic oscillations and substrate inhibition in enzymatic reactions. Its vector field reads
\begin{equation}
	\bm F_{\rm S}(x,y)
	=
	\begin{pmatrix}
		1-xy^{\gamma} \\
		\alpha y(xy^{(\gamma-1)}-1)
	\end{pmatrix},
\end{equation}
with parameters $\alpha=1.1$ and $\gamma=2$ chosen such that the system exhibits a stable limit cycle. The term $\gamma$ breaks the odd symmetry. 

\paragraph{Brusselator.}
The Brusselator is a canonical model of autocatalytic chemical reactions, originally introduced to describe oscillatory chemical kinetics (e.g., glycolytic reactions). Its vector field is
\begin{equation}
	\bm F_{\rm B}(x,y)
	=
	\begin{pmatrix}
		1-(b+1)x+c x^2 y \\
		b x-c x^2 y
	\end{pmatrix},
\end{equation}
with $b=2.5$ and $c=1.1$. The term $1$ breaks the odd symmetry.

\subsection{Coupling functions}
\label{sec:app:couplings}

We used twelve nonlinear coupling functions of the form
\begin{equation}
	\bm{G}(\bm{X}_j, \bm{X}_k, \bm{X}_l)
	=
	\begin{pmatrix}
		G^{(x)}(x_j,y_j,x_k,y_k,x_l,y_l) \\
		G^{(y)}(x_j,y_j,x_k,y_k,x_l,y_l)
	\end{pmatrix}.
\end{equation}

The twelve functions are then defined as 
\begin{alignat}{3}
\bm G_1 &: \quad G^{(x)} = x_j x_k x_l,      &\qquad& G^{(y)} = 0, \\
\bm G_2 &: \quad G^{(x)} = 0,                &\qquad& G^{(y)} = x_j x_k x_l, \\
\bm G_3 &: \quad G^{(x)} = y_j y_k y_l,      &\qquad& G^{(y)} = 0, \\
\bm G_4 &: \quad G^{(x)} = 0,                &\qquad& G^{(y)} = y_j y_k y_l, \\
\bm G_5 &: \quad G^{(x)} = x_k x_l^2,        &\qquad& G^{(y)} = 0, \\
\bm G_6 &: \quad G^{(x)} = 0,                &\qquad& G^{(y)} = x_k x_l^2, \\
\bm G_7 &: \quad G^{(x)} = y_k y_l^2,        &\qquad& G^{(y)} = 0, \\
\bm G_8 &: \quad G^{(x)} = 0,                &\qquad& G^{(y)} = y_k y_l^2, \\
\bm G_9 &: \quad G^{(x)} = x_k x_l,                &\qquad& G^{(y)} = 0, \\
\bm G_{10} &: \quad G^{(x)} = 0,                &\qquad& G^{(y)} = y_k y_l. \\
\bm G_{11} &: \quad G^{(x)} = y_k y_l,                &\qquad& G^{(y)} = 0. \\
\bm G_{12} &: \quad G^{(x)} = 0,                &\qquad& G^{(y)} = y_k y_l.
\end{alignat}
Functions $\bm{G}_1$-$\bm{G}_4$ are odd in each of the three arguments.
Functions $\bm{G}_5$-$\bm{G}_8$ are odd in $\bm{X}_k$ but even in $\bm{X}_j$ and $\bm{X}_l$.
Functions $\bm{G}_9$-$\bm{G}_{12}$ are odd in $\bm{X}_k$ and $\bm{X}_l$ but even in $\bm{X}_j$.

%


\section{Numerical results}
\label{sec:sm:numerics}

\subsection{Numerical phase reduction}

Numerically computing the phase reduced equation \eqref{eq:phase_reduction} in the main text requires computing the frequency $\omega$ and the interaction function $\Pi$, for which we need to compute the limit cycle solution $\bm{X}^c(\theta)$ and the phase sensitivity function $\bm{Z}(\theta)$ (Eq.~\eqref{eq:pi_definition} and Eq.~\eqref{eq:p_definition} in the main text).
First, the limit cycle and frequency are obtained through numerical integration of the evolution equation of an isolated oscillator, \cref{eq:original_system} in the main text, with $\kappa=0$. After a long transient, the system reaches the limit cycle, and we can record one period.
Second, the phase sensitivity function is the periodic solution to the adjoint equation
\begin{equation} \label{eq:adjoint}
    \omega \frac{d\bm{Z}(\theta)}{d\theta}=-J^\top (\theta)\bm{Z}(\theta)
\end{equation}
where $J^\top$ is the transpose of the Jacobian matrix of $\bm{F}$ evaluated on the limit cycle solution $\bm{X}^c(\theta)$. The phase sensitivity is then obtained by numerically integrating the adjoint equation backward, imposing the normalization condition $\bm{Z}(\theta)\cdot\frac{d\bm{X}^c(\theta)}{d\theta}=1$ \cite{nakao2016phase,ermentrout1996type}. 

With this numerical scheme, the phase sensitivity function can be obtained with high accuracy. The accuracy can always be increased by decreasing the integration time step and considering larger transient times to reach the periodic solution. The precision of the computed phase sensitivity function and limit cycle directly affect the precision of the phase interaction function Eq.~\eqref{eq:pi_definition} and its Fourier components \eqref{eq:pi_coefficients}. In this paper, we employ the Euler algorithm with a time step of $\delta t=10^{-4}$ and a transient of $1000$ periods to reach the periodic solution.

\subsection{Phase reduction across systems}

\begin{figure}[h]
    \centering
    \includegraphics[width=1\linewidth]{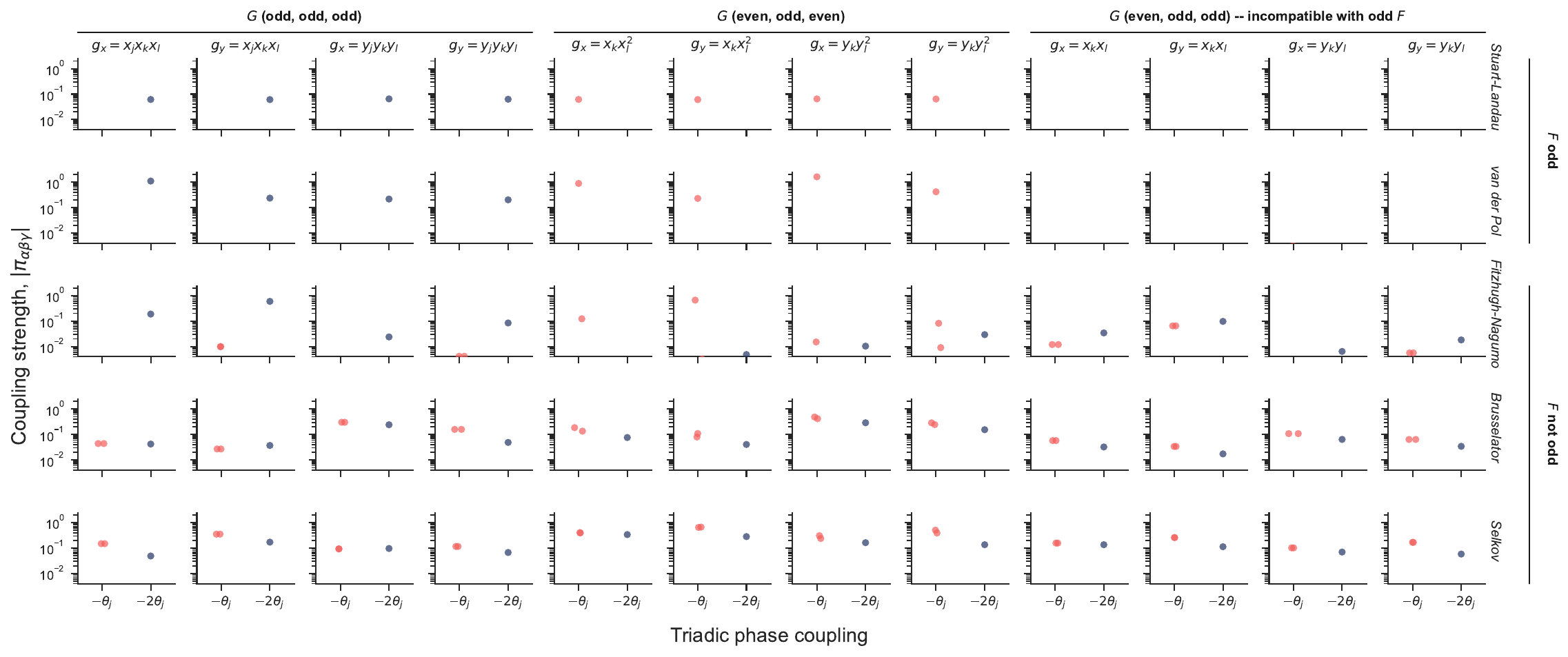}
    \caption{	\textbf{Symmetry of nonlinear oscillators suppresses classes of coupling in phase reduced models.}
    	We show the phase coupling coefficients $|\pi_{-2, 1, 1}|$ (blue), and $|\pi_{-1, 2, -1}|$ and $|\pi_{-1, -1, 2}|$ (red) for all five oscillators $\bm{F}$ and twelve coupling functions $\bm{G}$. This figure extends \cref{fig:numerical_classification}.}
    \label{fig:sm:triadic_coefficients}
\end{figure}

\Cref{fig:sm:triadic_coefficients} shows the numerically computed phase coupling coefficients $\pi_{-1, -1, 2}$, and $\pi_{-1,2,-1}$ (red) associated with the first-$j$ harmonic coupling functions $\sin(- \theta_j + \cdots )$, and $\pi_{-2, 1, 1}$ (blue) associated with second-$j$ harmonic $\sin(-2 \theta_j + \cdots )$, for all five nonlinear oscillators and twelve coupling functions. 
For odd $\bm{F}$, such as Stuart-Landau and van der Pol, the phase reduced models contain a single coupling function---$\sin(-2 \theta_j + \cdots )$ for functions $\bm{G}_1$-$\bm{G}_4$, but $\sin(- \theta_j + \cdots )$ for functions $\bm{G}_5$-$\bm{G}_8$---or none, for functions $\bm{G}_9$-$\bm{G}_{12}$. 
For non-odd $\bm{F}$, such as Selkov and the Brusselator, the phase reduced models contain a mix of coupling functions with first- and second-$j$ harmonics. 
FitzHugh-Nagumo has nearly odd $\bm{F}$ and consequently shows intermediate results where one type of coupling is much weaker than the other for $\bm{G}_1$-$\bm{G}_8$.
These results extend those shown \cref{fig:numerical_classification} and are all consistent with the symmetry rules detailed in the manuscript.

\subsection{Harmonic content of limit cycle and phase sensitivity function}

\begin{figure}[h]
    \centering
    \includegraphics[width=0.99\linewidth]{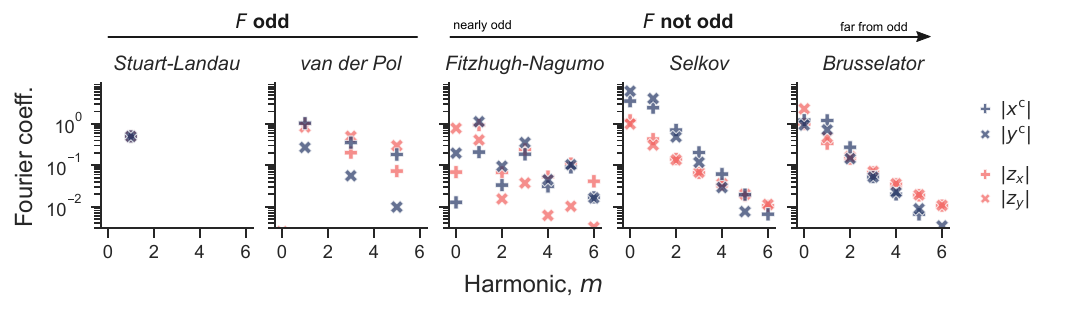}
    \caption{
    \textbf{Fourier coefficients of limit cycle and phase sensitivity function for all five oscillators.}
    Modulus of the Fourier coefficients of the $x$ ({\small$+$}) and 
    $y$ ({\small$\times$}) components of the limit cycle 
    $\bm{X}^c=(x,y)$ (red) and phase sensitivity function 
    $\bm{Z}=(Z_x,Z_y)$ (blue), for all five oscillators ordered from 
    odd $\bm{F}$ (left) to far from odd (right).
    Oscillators with odd $\bm{F}$---Stuart-Landau and van der Pol---contain only odd harmonics.
    FitzHugh-Nagumo is nearly odd, with even harmonics present but substantially weaker than odd ones ($\mathcal{S}=0.03$).
    Selkov and Brusselator have broken inversion symmetry and contain  harmonics of both parities ($\mathcal{S}=0.11$ and $0.13$).}
    \label{fig:sm:limit_cycle_coefficients}
\end{figure}

\Cref{fig:sm:limit_cycle_coefficients} shows the numerically computed coefficients of the limit cycle $\bm{X}^{\rm c}$ (red) and phase sensitivity function $\bm{Z}$ (blue), projected onto the $x$ ($+$) and $y$ ($\times$) variables, for all five nonlinear oscillators. 
Oscillators with odd $\bm{F}$ only have odd harmonics, such as Stuart-Landau or van der Pol, whereas those with non odd $\bm{F}$ have both odd and even harmonics---with weaker odd harmonics for nearly odd ones such as FitzHugh-Nagumo.

\subsection{Harmonic decay}

\begin{figure}[h]
    \centering
    \includegraphics[width=1\linewidth]{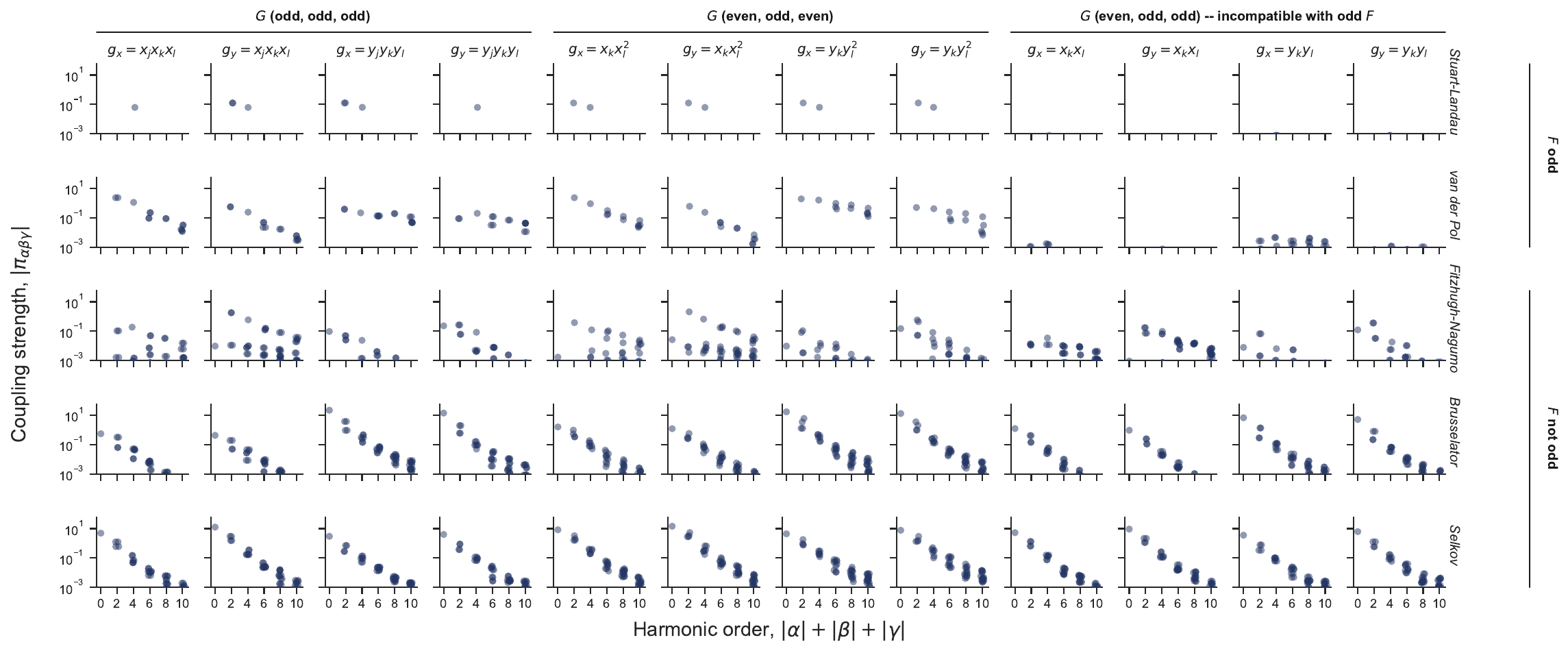}
    \caption{
    \textbf{Harmonic decay of three-body phase coupling coefficients.}
Modulus of the phase coupling coefficients $|\pi_{\alpha\beta\gamma}|$ 
as a function of harmonic order $|\alpha|+|\beta|+|\gamma|$, for all 
five oscillators (rows) and twelve coupling functions (columns), 
separated into three different parity combinations.
In all cases, coefficients decay rapidly with harmonic order, 
consistent with the smoothness of the underlying limit cycle 
(App.).
This justifies focusing on the lowest-order resonant terms 
$\sin(-2\theta_j+\theta_k+\theta_l)$ and 
$\sin(-\theta_j-\theta_k+2\theta_l)$, which dominate the 
phase dynamics.
}
    \label{fig:harmonic_decay}
\end{figure}

In Fig.~\cref{fig:harmonic_decay} we depict the modulus of the Fourier harmonics of the coupling coefficient versus the total harmonic order $|\alpha|+|\beta|+|\gamma|$. The figure clearly shows that the coefficient strengths decay exponentially with the total harmonic order. This justifies our special attention to the considered triplets $\pi_{-2,1,1}$, $\pi_{-1,-1,2}$ and $\pi_{-1,2,-1}$. 

\section{Analytical symmetry-based rules}
\label{sec:sm:symmetry_rules}

\subsection{Harmonic content of $p$}
\label{sec:sm:p_harmonics}


We start with an oscillator with an odd vector field $\bm{F}(-\bm{X}) = -\bm{F}(\bm{X})$. 
Then the limit cycle satisfies the following $\pi$-shift symmetry:
\begin{equation} \label{eq:sm:pi-shift}
\bm{X}^{\rm c}(\theta+\pi) = -\bm{X}^{\rm c}(\theta).
\end{equation}
Now, as in the main text, we define
\begin{equation}
\bm{P}(\theta_j,\theta_k,\theta_l)
=
\bm{G}\big(\bm{X}^{\rm c}(\theta_j),\bm{X}^{\rm c}(\theta_k),\bm{X}^{\rm c}(\theta_l)\big),
\end{equation}
which we can expand in Fourier modes
\begin{equation}
\bm{P}(\theta_j,\theta_k,\theta_l)
=
\sum_{\alpha,\beta,\gamma \in \mathbb{Z}}
\bm{p}_{\alpha,\beta,\gamma}
e^{i(\alpha\theta_j+\beta\theta_k+\gamma\theta_l)}.
\end{equation}


Suppose that $\bm{G}$ is even or odd in $\bm{X}_j$, i.e.
\begin{equation}
\bm{G}(-\bm{X}_j,\bm{X}_k,\bm{X}_l)
=
\sigma_j \bm{G}(\bm{X}_j,\bm{X}_k,\bm{X}_l),
\qquad
\sigma_j \in \{+1,-1\}.
\end{equation}
Using \cref{eq:sm:pi-shift}, we obtain
\begin{equation}
\bm{P}(\theta_j+\pi,\theta_k,\theta_l)
=
\sigma_j
\bm{P}(\theta_j,\theta_k,\theta_l) ,
\end{equation}
and from the Fourier expansion,
\begin{equation}
\bm{P}(\theta_j+\pi,\theta_k,\theta_l)
=
\sum_{\alpha,\beta,\gamma}
(-1)^\alpha
\bm{p}_{\alpha,\beta,\gamma}
e^{i(\alpha\theta_j+\beta\theta_k+\gamma\theta_l)}.
\end{equation}
Together, the last two equations imply the following constraint on the Fourier coefficients,
\begin{equation}
\sigma_j \bm{p}_{\alpha,\beta,\gamma}
=
(-1)^\alpha \bm{p}_{\alpha,\beta,\gamma} .
\end{equation}
Hence 
\begin{equation}
\bm{p}_{\alpha,\beta,\gamma} = 0
\quad
\text{for all } \alpha \text{ such that } (-1)^\alpha \neq \sigma_j.
\end{equation}
In other words, the parity symmetries of $\bm{G}$ are preserved in $\bm{P}$, for Class I oscillators:
\begin{itemize}
\item if $\bm{G}$ is even in $\bm{X}_j$ ($\sigma_j=+1$), then $\bm{p}_{\alpha,\beta,\gamma}=0$ for all odd $\alpha$;
\item if $\bm{G}$ is odd in $\bm{X}_j$ ($\sigma_j=-1$), then $\bm{p}_{\alpha,\beta,\gamma}=0$ for all even $\alpha$;
\item if $\bm{G}$ is neither odd nor even in $\bm{X}_j$, then no $\bm{p}_{\alpha,\beta,\gamma}$ need to vanish.
\end{itemize}
The same argument applies independently to $\bm{X}_k$ and $\bm{X}_l$.

\paragraph{Example 1: $\bm{G}_1 = x_j x_k x_l$.}

Here $\bm{G}_1$ is odd in each argument, so
\begin{equation}
\sigma_j=\sigma_k=\sigma_l=-1,
\end{equation}
and hence,
\begin{equation}
\bm{p}_{\alpha,\beta,\gamma} \neq 0
\quad \text{only if} \quad
\alpha,\beta,\gamma \text{ are all odd}.
\end{equation}

\paragraph{Example 2: $\bm{G}_2 = x_j^2 x_k x_l$.}

Here $\bm{G}_2$ is even in $\bm{X}_j$ and odd in $\bm{X}_k,\bm{X}_l$, so
\begin{equation}
\sigma_j=+1,
\qquad
\sigma_k=-1,
\qquad
\sigma_l=-1.
\end{equation}
Therefore
\begin{equation}
\bm{p}_{\alpha,\beta,\gamma} \neq 0
\quad \text{only if} \quad
\alpha \text{ even},
\quad
\beta,\gamma \text{ odd}.
\end{equation}

\paragraph{General statement.}

If $\bm{F}$ is odd, the limit cycle induces a $\pi$-shift symmetry in each phase variable.  
For each argument $\bm{X}_m$:

\begin{itemize}
\item if $\bm{G}$ is even in $\bm{X}_m$, only even Fourier coefficients in $\theta_m$ survive in $\bm{P}$;
\item if $\bm{G}$ is odd in $\bm{X}_m$, only odd Fourier coefficients in $\theta_m$ survive in $\bm{P}$;
\item if $\bm{G}$ has no definite parity in $\bm{X}_m$, no symmetry-enforced cancellation occurs in $\bm{P}$.
\end{itemize}

These vanishing conditions follow purely from symmetry. Conversely, if $\bm{F}$ is not odd, no symmetry-enforced cancellation of $\bm{P}$ coefficients occur either.

\subsection{Harmonic content of $\Pi$}
\label{sec:sm:pi_harmonics}

We now determine how the symmetry of the oscillator vector field $\bm{F}$ and of the interaction function $\bm{G}$ constrains the harmonic content of the reduced phase interaction $\bm{\Pi}$. The phase interaction coefficients are given in the main text as
\begin{equation}
\pi_{\alpha \beta \gamma}
=
\sum_{\substack{m\\\alpha+\beta+\gamma=0}}
\bm{z}_{\alpha-m} \cdot \bm{p}_{m\beta \gamma}.
\label{eq:sm_pi_coeff}
\end{equation}
From \cref{eq:sm_pi_coeff}, we see that $m$, $\beta$, and $\gamma$ are Fourier coefficients inherited from $\bm{P}$ and hence inherit parity constraints from both $\bm{G}$ and $\bm{F}$, as seen in the previous section. 
In addition, $\alpha-m$ is a Fourier coefficient inherited from $\bm{Z}$, but it is constrained by $m$ so that the parity of $\alpha$ is also determined jointly by the parity of Fourier modes of $\bm{F}$ and $\bm{G}$. Finally, the resonance condition 
\begin{equation}\label{eq:sm:resonance_condition}
    \alpha+\beta+\gamma=0
\end{equation}
further constrains $\beta$ and $\gamma$ and their parity: either one or all the coefficients must be even---e.g. (even , odd, odd) or (even, even, even).

\subsubsection{Odd $F$}

If the vector field is odd $\bm{F}(-\bm{X})=-\bm{F}(\bm{X})$,
then $\bm{Z}$ contains only odd Fourier harmonics. This, in turn, implies in \cref{eq:sm_pi_coeff} that 
\begin{equation}
\alpha-m \;\text{is odd}, 
\label{eq:sm_Z_odd}
\end{equation}
because $\bm{z}_{\alpha-m}=0$ for all even $\alpha-m$. 

All further constraints arise from the symmetries of $\bm{G}$.
In fact, the parity of index $m$ is entirely determined by the parity of $\bm{G}$ with respect to $\bm{X}_j$.
Indeed, we know from the previous section that if $\bm{G}$ is odd (even) with respect to $\bm{X}_j$, then, $m$ must be odd (even), because $\bm{p}_{m\beta \gamma} = 0$ for all even (odd) $m$. So, $m$ inherits its parity from $\bm{G}$, and if  $\bm{G}$ has not parity, all $m$ are allowed. 

Combining this constraint with \cref{eq:sm_Z_odd}, we see that the parity of $\bm{G}$ in $\bm{X}_j$ fully determines the parity of $\alpha$ in $\pi_{\alpha \beta \gamma}$. Indeed, 
\begin{itemize}
    \item if $\bm{G}$ is odd in $\bm{X}_j$: $m$ is odd and hence $\alpha$ is even. The resonance condition further imposes that $\beta + \gamma$ is even too.
    \item if $\bm{G}$ is even in $\bm{X}_j$: similarly, $m$ is even, and thus $\alpha$ is odd and $\beta + \gamma$ is also odd. 
    \item if $\bm{G}$ has no parity in $\bm{X}_j$: $m$ is not constrained, and thus $\alpha$, $\beta$, $\gamma$ can have any parity as long as their sum respects the resonance condition. 
\end{itemize}
As seen in the previous section, the parity of $\bm{G}$ in $\bm{X}_k$ and $\bm{X}_l$ imposes additional constraints on the parity of $\beta$ and $\gamma$. Indeed, oddness (evenness) in any of the two parameters induces oddness (evenness) in the associated coefficient. Note that the constraints $\alpha$, $\beta$, and $\gamma$, inherited from $\bm{G}$ and from the resonance condition, can be incompatible. In this case, the coefficient $\pi_{\alpha \beta \gamma}=0$ because the associated phase coupling function is not allowed. 

\begin{table}[h]
	\centering
	\caption{
    \textbf{Three-body interactions with odd $\bm{F}$:} parity combinations of $\bm{G}$ and corresponding parity constraints on the phase coupling functions, with lowest-harmonic examples. ``/" indicates that no phase couplings are allowed because the constraints on $\alpha$, $\beta$, and $\gamma$ are incompatible with the resonance condition \cref{eq:resonance}. \Cref{tab:combinations_odd_f_no_parity} extends this Table to $\bm{G}$ with no parity in some of its arguments.}
	\label{tab:combinations_odd_f}
	\begin{tabular}{lllr|lllc}
		\toprule
		$G_j$ & $G_k$ & $G_l$ & Ex. $\bm{G}$ & $\alpha$ & $\beta$ & $\gamma$ & Lowest-harmonic example \\
		\midrule
		odd  & odd  & odd  & $\bm{X}_j\bm{X}_k\bm{X}_l$ & even & odd  & odd  & $\sin(-2\theta_j + \theta_k + \theta_l)$ \\
		& even & even & $\bm{X}_j\bm{X}_k^2\bm{X}_l^2$ & & even & even & $\sin(-4\theta_j + 2\theta_k + 2\theta_l)$ \\
		& even & odd & $\bm{X}_j\bm{X}_k^2\bm{X}_l$ & & even & odd & / \\
		& odd & even & $\bm{X}_j\bm{X}_k\bm{X}_l^2$ & & odd & even & / \\
		even & even & odd  & $\bm{X}_k^2\bm{X}_l$ & odd  & even & odd  & $\sin(-\theta_j + 2\theta_k - \theta_l)$ \\
		& odd  & even &  $\bm{X}_k\bm{X}_l^2$ & & odd  & even & $\sin(-\theta_j - \theta_k + 2\theta_l)$ \\  
		& even  & even & $\bm{X}_k^2\bm{X}_l^2$ &  & even  & even & / \\
		& odd  & odd & $\bm{X}_k\bm{X}_l$ &  & odd  & odd & / \\
		\bottomrule
	\end{tabular}
\end{table}

All combinations are listed in \cref{tab:combinations_odd_f,tab:combinations_odd_f_no_parity} and we illustrate these rules with a few examples. In these example to avoid more cumbersome notation by $\bm{X}_j \bm{X}_k$ we mean the component product, not the dot product.
\begin{itemize}
    \item First, take $\bm{G} = \bm{X}_j \bm{X}_k \bm{X}_l$ which  is odd in all three arguments. We must thus have even $\alpha$, but odd $\beta$ and $\gamma$. The lowest-harmonic three-body interaction that satisfies this and the resonance condition is $\sin(-2\theta_j+\theta_k+\theta_l)$, which is the symmetric interaction considered in \cref{fig:numerical_classification} of the main text. Other higher-harmonic interactions allowed include $\sin(-4\theta_j+3\theta_k+\theta_l)$. Note that the asymmetric interaction $\sin(-\theta_j-\theta_k+2\theta_l)$ is not allowed by this choice of $\bm{G}$. Any $\bm{G}$ that is polynomial where all three arguments appear with odd powers will allow the same phase interactions. 

    \item Second, consider $\bm{G} = \bm{X}_k \bm{X}_l^2$, which is even in $\bm{X}_j$ and $\bm{X}_l$ but odd in $\bm{X}_k$. This implies that we must have odd $\alpha$ and $\beta$ but even $\gamma$. The lowest-harmonic three-body interaction that satisfies this and the resonance condition is $\sin(-\theta_j-\theta_k+2\theta_l)$, which is the asymmetric interaction considered in \cref{fig:numerical_classification} of the main text. Other higher-harmonic interactions allowed include $\sin(-3\theta_j-\theta_k+4\theta_l)$. Note that the symmetric interaction $\sin(-2\theta_j+\theta_k+\theta_l)$ is not allowed by this choice of $\bm{G}$. 

    \item Third, consider $\bm{G} =\bm{X}_k^2 \bm{X}_l^2$, which is even in all three arguments. This implies that we must have odd $\alpha$ but even $\beta$ and $\gamma$. This is incompatible with the resonance condition and all $\pi_{\alpha \beta \gamma}=0$: no phase interaction is allowed for this choice of $\bm{G}$ for odd oscillators. 

    \item Finally, take $\bm{G} = \bm{X}_j \bm{X}_k \bm{X}_l + \bm{X}_k \bm{X}_l^2+ \bm{X}_k^2 \bm{X}_l^2 $, which has no parity in any of its arguments. There are no constraints on the coefficients, and all $\sin(\alpha \theta_j + \beta \theta_k + \gamma \theta_l)$ that satisfy the resonance condition are allowed. Any polynomial $\bm{G}$ where each argument appears with both odd and even powers allows the same phase interactions. Nevertheless, because phase reduction is linear in $\bm{G}$, you can split it into three parts, $\bm{G}_1=\bm{X}_j \bm{X}_k \bm{X}_l$, $\bm{G}_2= \bm{X}_k \bm{X}_l^2$ and $\bm{G}_3=\bm{X}_j^2 \bm{X}_k^2 \bm{X}_l^2$ and analyze them independently. Since we have analyzed them previously, we know which terms provide each phase interaction, namely $\sin(-2\theta_j+\theta_k+\theta_l)$ from $\bm{G}_1$ and $\sin(-\theta_j-\theta_k+2\theta_l)$ from $\bm{G}_2$.
\end{itemize}

Overall, for oscillators with odd $\bm{F}$, the parity of $\bm{G}$ in each of its arguments determines which three-body phase interactions are allowed and which are not: all $\sin(\alpha \theta_j + \beta \theta_k + \gamma \theta_l)$, none, or some. 
Importantly, all phase interactions are allowed only if $\bm{G}$ has no parity in $\bm{X}_j$, which is not very common in the literature, because it requires a polynomial where $\bm{X}_j$ appears with both odd and even powers. Conversely, as soon as $\bm{G}$ has parity in $\bm{X}_j$, either first- or second-$j$ harmonic is allowed.



\subsubsection{Not odd $F$}

If $\bm{F}$ is not odd, the limit cycle and $\bm{Z}$ contain both even and odd harmonics. Then $\alpha-m$ in Eq.~\eqref{eq:sm_pi_coeff} can be even or odd, and the above parity constraints no longer apply. Any phase combination satisfying \cref{eq:sm:resonance_condition} may in principle appear, independently of the symmetries of $\bm{G}$. These oscillators therefore generically generate both symmetric and asymmetric many-body phase interactions.

\begin{table}[h]
	\centering
	\caption{
    \textbf{Three-body interactions with odd $\bm{F}$:} parity combinations of $\bm{G}$---when $\bm{G}$ has not parity in two of its arguments---and corresponding parity constraints on the phase coupling functions, with lowest-harmonic examples. ``/" indicates that $\bm{G}$ has no parity in the corresponding argument. ``*'' indicates that the corresponding coefficient has no parity constraint: it can be even or odd.
    Note that if $\bm{G}$ has no parity in only one argument, the rules of \cref{tab:combinations_odd_f} hold because of the resonance conditions. If it has no parity in any of its arguments, then all three-body phase couplings are allowed. Note that absence of parity in $\bm{G}$ is only possible if it is non-polynomial.}
    \label{tab:combinations_odd_f_no_parity}
	\begin{tabular}{lll|lllcc}
		\toprule
		$G_j$ & $G_k$ & $G_l$ & $\alpha$ & $\beta$ & $\gamma$ & Lowest-harmonic example \\
		\midrule
		odd  & /  & /  &  even & *  & *  & $\sin(-2\theta_j + \theta_k + \theta_l) + \sin(-4\theta_j + 2\theta_k + 2\theta_l)$  \\
		even & / & / & odd & * & * & $\sin(-\theta_j + 2\theta_k - \theta_l) + \sin(-\theta_j - \theta_k + 2\theta_l)$  \\
		/ & odd & / & * & odd & * &  $\sin(-2\theta_j + \theta_k + \theta_l) + \sin(-\theta_j - \theta_k + 2\theta_l) $ \\
		/ & even & / & * & even & * & $\sin(-4\theta_j + 2\theta_k + 2\theta_l) + \sin(-\theta_j + 2\theta_k - \theta_l)$   \\
		/ & / & odd  & *  & * & odd  & $\sin(-2\theta_j + \theta_k + \theta_l) + \sin(-\theta_j + 2\theta_k - \theta_l)$  \\
		/ & /  & even & * & * & even & $\sin(-4\theta_j + 2\theta_k + 2\theta_l) + \sin(-\theta_j - \theta_k + 2\theta_l)$   \\  
		\bottomrule
	\end{tabular}
\end{table}

\subsection{Generalization to $n$-body interactions}
\label{sec:app:nbody}

The symmetry rules derived in the main text for three-body interactions 
extend directly to $n$-body interactions of any size.
Consider $N$ oscillators with $n$-body coupling,
\begin{equation}
    \dot{\bm{X}}_j = \bm{F}(\bm{X}_j) + \kappa \sum_{\bm{k}} 
    A_{j\bm{k}}\,\bm{G}(\bm{X}_j, \bm{X}_{\bm{k}}),
\end{equation}
where $\bm{k}=(k_1,\ldots,k_{n-1})$ and $\bm{G}:\mathbb{R}^{nd}\to\mathbb{R}^d$.
Phase reduction yields
\begin{equation}
    \dot\theta_j = \omega + \kappa\sum_{\bm{k}} A_{j\bm{k}}\,
    \Pi(\theta_j,\theta_{\bm{k}}) + O(\kappa^2),
\end{equation}
with the $n$-body phase coupling function
\begin{equation}
    \Pi(\theta_j,\theta_{\bm{k}}) = \frac{1}{2\pi}\int_0^{2\pi}
    \bm{Z}(\theta_j+\varphi)\cdot
    \bm{P}(\theta_j+\varphi,\theta_{\bm{k}}+\varphi)\,d\varphi,
\end{equation}
where $\bm{p}(\theta_j,\theta_{\bm{k}}) = 
\bm{G}(\bm{X}^c(\theta_j),\bm{X}^c(\theta_{k_1}),\ldots,\bm{X}^c(\theta_{k_{n-1}}))$.
Expanding in Fourier modes,
\begin{equation}
    \Pi(\theta_j,\theta_{\bm{k}}) = 
    \sum_{\substack{\alpha,\alpha_1,\ldots,\alpha_{n-1}\\
    \alpha+\sum_i\alpha_i=0}}
    \pi_{\alpha,\alpha_1,\ldots,\alpha_{n-1}}\,
    e^{i(\alpha\theta_j+\sum_i\alpha_i\theta_{k_i})},
\end{equation}
the resonance condition $\alpha+\sum_{i=1}^{n-1}\alpha_i=0$ 
follows from the averaging integral, as in the three-body case.
The Fourier coefficients are given by the convolution
\begin{equation}
    \pi_{\alpha,\alpha_1,\ldots,\alpha_{n-1}} = 
    \sum_{\substack{m\\\alpha+\sum_i\alpha_i=0}}
    \bm{z}_{\alpha-m}\cdot\bm{p}_{m,\alpha_1,\ldots,\alpha_{n-1}},
    \label{eq:pi_nbody}
\end{equation}
which has the same structure as Eq.~(\ref{eq:pi_coefficients}) in the 
main text, with $\alpha-m$ inherited from $\bm{Z}$ and 
$m,\alpha_1,\ldots,\alpha_{n-1}$ inherited from $\bm{p}$.

The parity constraints follow identically.
If $\bm{F}$ is odd, then $\bm{Z}$ contains only odd harmonics, 
so $\alpha-m$ is always odd in Eq.~(\ref{eq:pi_nbody}).
The parity of $m$ is fixed by the parity of $\bm{G}$ in $\bm{X}_j$, 
which in turn fixes the parity of $\alpha$: odd (even) $\bm{G}$ in 
$\bm{X}_j$ forces even (odd) $\alpha$.
The parity of each $\alpha_i$ is independently fixed by the parity 
of $\bm{G}$ in $\bm{X}_{k_i}$.
The resonance condition $\alpha+\sum_i\alpha_i=0$ then acts as a 
compatibility check: parity assignments that satisfy it yield allowed 
couplings; incompatible assignments give $\pi_{\alpha,\alpha_1,\ldots,\alpha_{n-1}}=0$.
If $\bm{F}$ is not odd, all constraints are lifted and all 
harmonics satisfying the resonance condition are generically nonzero.

These constraints are summarized for $n=4$ in Table~\ref{tab:parity_combinations_4body}, 
Note that for larger $n$, more parity combinations are compatible with 
the resonance condition, admitting a richer set of allowed couplings---but the core selection mechanism is unchanged.

\begin{table}[h]
	\centering
	\caption{
    \textbf{Four-body interactions with odd $\bm{F}$:} parity combinations of $\bm{G}$ and corresponding parity constraints on the phase coupling functions, with lowest-harmonic examples. We only show the non-vanishing combinations---that is, compatible with the resonance condition \cref{eq:resonance}---for $\bm{G}$ with parity in all of its arguments.
    }
	\label{tab:parity_combinations_4body}
	\begin{tabular}{lllll|llllc}
		\toprule
		$G_j$ & $G_k$ & $G_l$ & $G_m$ & Ex. $\bm{G}$ & $\alpha$ & $\beta$ & $\gamma$ & $\delta$ & Lowest-harmonic example \\
		\midrule
		odd  & odd  & odd  & odd  & $\bm{X}_j\bm{X}_k\bm{X}_l\bm{X}_m$       & even & odd  & odd  & even & $\sin(-2\theta_j+\theta_k+\theta_l+2\theta_m)$ \\

		  & odd  & even & even & $\bm{X}_j\bm{X}_k\bm{X}_l^2\bm{X}_m^2$   &  & even & even & even & $\sin(-2\theta_j-2\theta_k+2\theta_l+2\theta_m)$  \\
		\midrule
		even & odd  & odd  & odd  & $\bm{X}_k \bm{X}_l \bm{X}_m$         & odd  & odd  & odd  & odd  & $\sin(-\theta_j+\theta_k-\theta_l+\theta_m)$ \\
		 & odd  & even & even & $\bm{X}_k \bm{X}_l^2 \bm{X}_m^2$     &   & odd  & even & even & $\sin(-\theta_j+\theta_k-2\theta_l+2\theta_m)$  \\
		\bottomrule
	\end{tabular}
\end{table}

\section{Low harmonic derivation}
\label{sec:sm:low-harmonics}

Here, we provide the full derivation and complete tables of 
phase coupling coefficients obtained from the low-harmonic 
approximation described in the appendix 
Analytical expressions were derived and verified using a 
symbolic computation script, available in the code 
repository~\url{https://github.com/maximelucas/phase_reduction}.

We truncate the limit cycle and phase sensitivity function at the second harmonic
\begin{align}
	X(\theta) &= x_0 + x_1 e^{i\theta} + x_2 e^{2i\theta} 
	+ \text{c.c.}, \label{eq:sm:xc_trunc}\\
	Z(\theta) &= z_0 + z_1 e^{i\theta} + z_2 e^{2i\theta} 
	+ \text{c.c.}, \label{eq:sm:z_trunc}
\end{align}
justified by harmonic decay (see App.). We treat both as scalars for clarity. This is a strong restriction that allows to proceed with the theory without the need to discuss the possibilities of getting zeros because of orthogonal vectors. In general there will not be orthogonalities so the present results are robust.
Because of harmonic decay, here we assume the coefficients to decay linearly: $x_k\sim \epsilon^k$ and $z_k\sim \epsilon^k$, with $\epsilon\ll1$.  We retain all 
terms up to order $\epsilon^4$; the first neglected contribution 
arises from the third harmonic at order $\epsilon^6$.
Substituting \cref{eq:sm:xc_trunc,eq:sm:z_trunc} into 
\cref{eq:pi_coefficients} and retaining only resonant terms 
($\alpha+\beta+\gamma=0$) yields all nonzero $\pi_{\alpha\beta\gamma}$ 
to this order.

Following the main text, we consider the two interaction functions $G(x_j, x_k, x_l)=x_j x_k x_l$ and $G(x_j, x_k, x_l)=x_k x_l^2$.
\Cref{tab:sm:pi_xjxkxl,tab:sm:pi_xkxl2} give
the full set of nonzero $\pi_{\alpha\beta\gamma}$ at order
$\epsilon^4$.  Since $\pi_{\alpha \beta \gamma}=\pi_{-j-k-l}^*$, we only write the ones with positive $j$, if $j=0$ we only write the ones with negative $k$. Several features are worth noting:
\begin{enumerate}
	\item The symmetry rules
	are confirmed throughout: for odd $\bm{F}$, all coefficients
	involving even harmonics ($x_0$, $x_2$, $z_0$, $z_2$) vanish,
	leaving only the predicted nonzero terms.
	
	\item \textit{Pairwise-like terms} $\pi_{-1,1,0}$ and $\pi_{-1,0,1}$---corresponding to $\sin(\theta_j-\theta_k)$ and
	$\sin(\theta_j-\theta_l)$---appear at order $\epsilon^2$
	for non-odd $\bm{F}$ in both cases, but for
	$\bm{G}=x_kx_l^2$ they are present regardless of
	$x_0$ since $\bm{G}$ does not depend on $x_j$. 
	
	\item \textit{Forcing-like terms} $\pi_{0,\beta,\gamma}$ such as
	$\sin(\theta_k-\theta_l)$ are present even for odd $\bm{F}$
	and represent collective frequency corrections that do not
	affect synchronization for identical oscillators.
	
	\item For $\bm{G}=x_jx_kx_l$ the equality
	$\pi_{-1,-1,2}=\pi_{-1,2,-1}$ ensures the full phase model
	preserves the permutation symmetry of $\bm{G}$, whereas for
	$\bm{G}=x_kx_l^2$ these differ since $\bm{G}$ is
	asymmetric under $k\leftrightarrow l$.
\end{enumerate}

\begin{table}[h]
    \centering
    \begin{tabular}{rrr|l}
    	\toprule
$\alpha$ & $\beta$ & $\gamma$ & \textbf{Expression for $G(x_j, x_k, x_l)=x_j x_k x_l$} \\
\midrule
0 & 0 & 0 & $x_0^3 z_0 + (x_0^2 x_1^{*} z_1 + x_0^2 x_1 z_1^{*}) \epsilon^2 + (x_0^2 x_2^{*} z_2 + x_0^2 x_2 z_2^{*}) \epsilon^4$ \\
0 & -1 & 1 & $x_0 x_1 x_1^{*} z_0 \epsilon^2 + (x_1 x_1^{*2} z_1 + x_1^2 x_1^{*} z_1^{*}) \epsilon^4$ \\
0 & -2 & 2 & $x_0 x_2 x_2^{*} z_0 \epsilon^4$ \\
-1 & 2 & -1 & $(x_1^{*2} x_2 z_0 + x_0 x_1^{*} x_2 z_1^{*}) \epsilon^4$ \\
-1& 1 & 0 & $(x_0 x_1 x_1^{*} z_0 + x_0^2 x_1 z_1^{*}) \epsilon^2 + (x_0 x_1 x_2^{*} z_1 + x_0 x_1^2 z_2^{*}) \epsilon^4$ \\
-1& 0 & 1 & $(x_0 x_1 x_1^{*} z_0 + x_0^2 x_1 z_1^{*}) \epsilon^2 + (x_0 x_1 x_2^{*} z_1 + x_0 x_1^2 z_2^{*}) \epsilon^4$ \\
-1& -1 & 2 & $(x_1^{*2} x_2 z_0 + x_0 x_1^{*} x_2 z_1^{*}) \epsilon^4$ \\
-2& 2 & 0 & $(x_0 x_2 x_2^{*} z_0 + x_0 x_1^{*} x_2 z_1^{*} + x_0^2 x_2 z_2^{*}) \epsilon^4$ \\
-2& 1 & 1 & $(x_1^2 x_2^{*} z_0 + x_1^2 x_1^{*} z_1^{*} + x_0 x_1^2 z_2^{*}) \epsilon^4$ \\
-2& 0 & 2 & $(x_0 x_2 x_2^{*} z_0 + x_0 x_1^{*} x_2 z_1^{*} + x_0^2 x_2 z_2^{*}) \epsilon^4$ \\
\bottomrule
\end{tabular}
    \caption{\textbf{Low-harmonic analytical expressions.}
    	Phase coupling coefficients $\pi_{\alpha\beta\gamma}$
    	at order $\epsilon^4$ for $\bm{G}=(x_jx_kx_l,0)$, odd in all
    	arguments. For odd $\bm{F}$, only $\pi_{-2,1,1}$ survives;
    	all other coefficients vanish since they involve even harmonics
    	$x_0$, $x_2$, $z_0$, or $z_2$.}
    \label{tab:sm:pi_xjxkxl}
\end{table}

\begin{table}[h]
\centering
\begin{tabular}{rrr | l}
\toprule
\textbf{$j$} & \textbf{$k$} & \textbf{$l$} & \textbf{Expression for $G(x_j, x_k, x_l)=x_k x_l^2$} \\
\midrule
0 & 0 & 0 & $x_0^3 z_0 + 2 x_0 x_1 x_1^{*} z_0 \epsilon^2 + 2 x_0 x_2 x_2^{*} z_0 \epsilon^4$ \\
0 & -1 & 1 & $2 x_0 x_1 x_1^{*} z_0 \epsilon^2 + 2 x_1^{*2} x_2 z_0 \epsilon^4$ \\
0 & -2 & 2 & $(x_1^2 x_2^{*} z_0 + 2 x_0 x_2 x_2^{*} z_0) \epsilon^4$ \\
-1 & 2 & -1 & $2 x_0 x_1^{*} x_2 z_1^{*} \epsilon^4$ \\
-1 & 1 & 0 & $x_0^2 x_1 z_1^{*} \epsilon^2 + 2 x_1^2 x_1^{*} z_1^{*} \epsilon^4$ \\
-1 & 0 & 1 & $2 x_0^2 x_1 z_1^{*} \epsilon^2 + 2 x_0 x_1^{*} x_2 z_1^{*} \epsilon^4$ \\
-1 & -1 & 2 & $(x_1^2 x_1^{*} z_1^{*} + 2 x_0 x_1^{*} x_2 z_1^{*}) \epsilon^4$ \\
-2 & 2 & 0 & $x_0^2 x_2 z_2^{*} \epsilon^4$ \\
-2 & 1 & 1 & $2 x_0 x_1^2 z_2^{*} \epsilon^4$ \\
-2 & 0 & 2 & $(x_0 x_1^2 z_2^{*} + 2 x_0^2 x_2 z_2^{*}) \epsilon^4$ \\
\bottomrule
\end{tabular}
\caption{
	\textbf{Low-harmonic analytical expressions.}
	Phase coupling coefficients $\pi_{\alpha\beta\gamma}$
	at order $\epsilon^4$ for $\bm{G}=(x_kx_l^2,0)$, odd in
	$\bm{X}_k$ and even in $\bm{X}_j$ and $\bm{X}_l$. For odd $\bm{F}$, only $\pi_{-1,-1,2}$ survives. Pairwise-like
	terms appear at order $\epsilon^2$ regardless of $x_0$ since
	$\bm{G}$ is independent of $\bm{X}_j$; and
	$\pi_{-1,-1,2}\neq\pi_{-1,2,-1}$ since $\bm{G}$ is asymmetric
	under $k\leftrightarrow l$.}
\label{tab:sm:pi_xkxl2}
\end{table}

%

\clearpage

\end{document}